\documentclass[11pt, a4paper]{article}
\usepackage{vmargin,color} 
\setmarginsrb{2.0cm}{2.5cm}{2.0cm}{1.0cm}{0.0cm}{0.0cm}{0.0cm}{1.0cm} 
\usepackage[french,english]{babel} 
\usepackage[T1]{fontenc} 
\usepackage{cite,graphicx,float,setspace,amssymb,amsmath} 
\usepackage{graphicx} 

\newcommand{\equa}[1]{\begin{eqnarray} \label{#1}} 
\newcommand{\auqe}{\end{eqnarray}} 
\newcommand{\tab}[1]{\begin{tabular}{#1}} 
\newcommand{\bat}{\end{tabular} \\ }

 \newcommand{\be}{\beta} \newcommand{\la}{\lambda}
\newcommand{\ga}{\gamma} \newcommand{\ep}{\epsilon} 
\newcommand{\de}{\delta} \newcommand{\gD}{\Delta}  

\providecommand{\abs}[1]{\left\vert#1\right\vert} 

 \newcommand{\s}{\sigma} \newcommand{\om}{\omega}
\newcommand{\Om}{\Omega}
\newcommand{\tend}{\rightarrow} 
\newcommand{\mt}[1]{\tilde{#1}} 

\newcommand{\fik}{\varphi_{k}(z)}
\newcommand{\fix}[1]{\varphi_{#1}(z)}

\newcommand{\szz}{S_{\gD z\;\gD z}}
\begin{document} 
\selectlanguage{english} 
\title{
{\bf A statistical field theory approach applied to the liquid vapor interface}
}
\author
{~V. Russier$^a$         \footnote{~russier@icmpe.cnrs.fr} 
~and ~J.-M. Caillol$^b$  \footnote{~jean-michel.caillol@th.u-psud.fr}  \\
$^a$~ICMPE, UMR 7182 CNRS and Universit\'e Paris Est \\
2-8 rue Henri Dunant, 94320 Thiais, France. \\
$^b$~Laboratoire de Physique Th\'eorique, UMR 8627  CNRS and 
Universit\'e de Paris Sud, \\ bat. 210, 91405 Orsay Cedex, France. 
}

\date{}
\maketitle
\abstract {
% Resume
%
Last years, there has been a renewed interest in the utilization of 
statistical field theory methods for the description of systems at equilibrium both in 
the vicinity and away from critical points, in particular in the field of liquid
state physics. These works deal in general with homogeneous systems, although 
recently the study of liquids in the vicinity of hard walls has also been considered
in this way. On the other hand, effective Hamiltonian pertaining to the $\phi^4$ 
theory family have been written and extensively used for the description 
of inhomogeneous systems either at the simple interface between equilibrium
phases or for the description of wetting.
In the present work, we focus on a field theoretical description of the liquid
vapor interface of simple fluids. We start from the representation of the grand partition
function obtained from the Hubbard-Stratonovich transform leading to an exact formulation
of the problem, namely neither introducing an effective Hamiltonian nor associating
the field to the one-body density of the liquid. Using as a reference system the
hard sphere fluid and imposing the coexistence condition, the expansion of the
Hamiltonian obtained yields a usual $\phi^4$ theory without unknown parameter.
An important point is that the so-called capillary wave theory appears as
an approximation of the one-loop theory in the functional expansion of
the Hamiltonian, without any need to an underlying phenomenology. 
\\
PACS: s61.30.Hn; 64.70.F; 68.03.Cd.  
}
\eject
%%%%%%%%%%%%%%%%%%%% 
\section{Introduction } 
\label{intro} 
%%%% Introduction apres cette ligne 
The structure in the interface between two fluid phases at
coexistence plays a central role in numerous specific situations,
as for instance wetting transitions and related phenomena
\cite{sull, lipo_1, dietrich_1, lipo_5, brez, parry_s, hrt},
and presents a great interest in itself from a theoretical point of view.
Interfaces between coexisting
subcritical phases at equilibrium
provides some of the key features of more complex situations.
For simple fluids the most important characteristic
is the presence of thermally activated capillary waves \cite{bls} and their
coupling with bulk fluctuations. Capillary waves are usualy described
through the introduction of an effective surface Hamiltonian written as a
functional of the interface height $h(\vec{s})$
($\vec{s}$: coordinate parallel to the interface) and {\it a priori} built
from a phenomenological description of the interface.
Then this Hamiltonian, $H[h(\vec{s})]$, can be treated in the framework
of field theoretical methods \cite{lipo_1, lipo_5}, or renormalization group
(RG) \cite{wet_rg_1} theory.
Recently, the hierarchical reference theory ($HRT$) has been generalized to
inhomogeneous case in order to deal with wetting transitions \cite{hrt}.

At the liquid vapor interface when $H[h(\vec{s})]$ is treated
at the gaussian level the capillary wave theory ($CWT$) is recovered,
leading to the well known $1/q^2$ behavior for the height-height structure
factor $\szz(q)$. Using a 3D $\phi^4$ model \cite{brez, parry_b, parry_s, brill_surf}
the theory has been extended
in the sense that the stochastic variable of the effective Hamiltonian
does not coincide any more with the height $h(\vec{s})$.
An alternative way to deduce an effective Hamiltonian from a DFT functional
including the effect of
the local curvature of the interface has been introduced in refs. 
\cite{eff_h_dietr, dft_md}.
Recently \cite{blokhuis}, a model for the density profile
based on an extension of a displaced profile type of approximation,
where the profile is written as a function of $(z-h(\vec{s}))$ was considered.
In addition, together with the surface fluctuations described by the height
$h(\vec{s}))$, the bulk phase fluctuations are added.

The purpose of the present work is to provide a simple description of
the liquid / vapor interface structure of simple fluids in the framework of
statistical field theory. We start from the Hubbard Stratonovich ($KSSHE$)
\cite{jmc_1} transform to get a representation of the grand-potential from which we
deduce the surface tension and the surface structure factor at the
the so-called one loop approximation level.
This is done {\it via} the determination
of the eigenvalues and eigenfunctions of the second functional derivative
of the mean field $KSSHE$ Hamiltonian.
Taking into account the whole spectrum of eigenstates is of the outmost
importance to obtain the correct result.
One of the salient points of this work is that one recovers both the $CWT$ and its
first extension, namely the appearance of the bending rigidity factor $k$ and
the coupling between surface and bulk correlations in $S(q)$,
without invoking any {\it ad hoc} phenomenology.

%
%%%% Section 1 apres cette ligne
\section {KSSHE transform and mean-field approximation}
\label{ksshe}
We briefly recall the important steps leading to the
approriate statistical field theory formulation of the problem;
the interested reader is refered to ref.\cite{jmc_1, siegert}
and \cite{brill_bulk} for an alternative formulation.
We consider a simple fluid whose pair interaction potential includes a hard
sphere repulsive part and a soft attractive part, denoted v(r). Let
\mbox{$w(r) = -\beta$v(r)} as usual and suppose that only purely attractive 
potentials are considered ($w$ is a positive definite operator, {\it i.e.} 
$\mt{w}(k)$ > 0). In the core region ($r < \s$), $w$ can be chosen at
will or regularized in order to fulfill a conveniently chosen criterion.
We work in the grand canonical ensemble, namely at constant chemical potential
\mbox{$\nu = \beta\mu$}, and we consider the $GC$ partition function related to
the grand potential \mbox{$\Xi = \exp(-\beta\Om)$}.
Using the Hubbard Stratonovich transformation
one can get $\Xi$ in the form of a functional integral
of $\exp(-H[\phi])$ over the field $\phi(\vec{r})$ which makes the link
with statistical field theory, $H[\phi]$ being the effective Hamiltonian
\equa{ksshe_1}
H = \frac{1}{2}<\phi|w^{-1}|\phi> - \ln(\Xi_{Hs}[\bar{\nu}+\phi]) ,
\auqe
with $\bar{\nu}$ = $\nu - w(0)/2$ and $\Xi_{HS}[\nu]$
is the hard sphere grand partition function at the local chemical potential 
$\nu(\vec{r})$. A Landau-Ginzburg form is obtained from
a functional Taylor expansion of $\ln(\Xi_{HS})$
arround some reference chemical potential of the hard sphere fluid, $\nu_0$.
Performing an expansion up to order $k^2$ of the propagator,
\mbox{$\mt{\gD}^{-1}(k) = K_0 + K_2k^2$}, we are left with
\begin{subequations}
\label{ksshe_3}
\equa{ksshe_3a}
H[\phi] &=& H_0 + \int
\left(\frac{K_2}{2} \left( \frac{\partial \phi}{\partial \vec{r}} \right)^2
+ \frac{K_0}{2} \phi^2 + V(\phi)
-  B \phi   \right) d\vec{r}  
\auqe
\equa{ksshe_3b}
V(\Phi) = \sum_{n \geq 3} \frac{u_n}{n!}\phi^n
\auqe
\equa{ksshe_3c}
u_n &=& - \frac{\partial ^n (\beta P_{hs})}{\partial \nu^n}(\nu = \nu_0)
 \equiv -(\beta P_{hs})^{(n)}[\nu_0]
\auqe
\end{subequations}
where in addition, we have neglected the $k^2$ dependence of the $u_n$.
This formulation allows an exact mapping between the densities and their
correlations
with the mean value and the correlations of the field \cite{jmc_1}.
For instance at coexistence, the densities $\rho_l$ and $\rho_g$ of
the two phases correspond to the two values $<\phi>_l$ and $<\phi>_l$.
The external field $B$ in (\ref{ksshe_3a}) is given by :
\equa{champ_ext}
B = w^{-1}*\gD \nu + \rho_{hs}\left[ \nu_0 \right] \makebox[1.5 cm]{with}
\gD\nu = \bar{\nu} - \nu_0.
\nonumber \auqe
The value of $\nu_0$ can be chosen in such a way that the $\phi^3$ term
vanishes and the coexistence condition between the
liquid and vapor phases is $B = 0$. $K_0$ is related to the deviation from the
critical temperature: $t = (T_c - T)/T_c~\propto~(-K_0)$.
The interaction part of the Hamiltonian is denoted by $V(\phi)$.

In the inhomogeneous system,
we assume a special realization of the two phases coexistence:
we impose explicitly the occurence of a bulk liquid and a bulk gas,
at densities $\rho_l$ and $\rho_g$ separated by a planar
surface located at \mbox{$z = z_0$}. The inhomogeneous mean field equation,
\mbox{$\de H[\phi]/\de\phi$ = 0},
leads to the mean field profile $\phi_c(z)$ = \mbox{$\phi_b\tanh(c(z - z_0))$},
when $V(\phi)$ is restricted to the order $\phi^4$. Here, $c=(-K_0/2K_2)^{1/2}$,
the inverse of the bulk correlation length,
is also the inverse of the intrinsic interface width and $\tend~0$
as $t^{1/2}$ when $t~\tend~0$. The mean field surface tension follows from
the identity $\Omega_{MF}[\nu]$ = $H[\phi_c(z)]$ with the result 
\equa{ga_mf}
\be\s^2\ga_{MF} = 4\s^2(-2K_2K_0^3/u_4^2)^{1/2},
\nonumber \auqe
which is similar to the expression given by Brilliantov \cite{brill_surf} with
however another definition for the reference system. Note that for $t~\tend~0$,
\mbox{$\ga_{MF} \sim t^{3/2}$}.
Finaly, for numerical calculations, we took for the $u_n$ their Carnahan
Starling values and the potential is regularized either in a W.C.A. scheme or from
ref. \cite{jmc_2}. 

%
%%%%  Section 2 apres cette ligne
\section {One loop equations}
\label{one_loop}
In order to go beyond the mean-field level we expand the Hamiltonian in the
vicinity of $\phi_c(z)$.
The first correction stems from the second order term and we have ($z_0 = 0$),
\equa{gauss_1}
H[\phi = \phi_{c} + \chi] &\simeq&
H[\phi_{c}] + \frac{1}{2}\int \chi(1) H^{(2)}(1,2) \chi(2) d1 d2
\auqe
Introducing the function \mbox{$g(z) = (K_0 + V''(\phi_{c}(z)))/K2$}, after a Fourier
transform parallel to the surface, the operator $H^{(2)}$ is diagonalized in the
set of the eigenfunctions ${\varphi_{n}}$ solution of the eigenvalues equation
($\ep_n = K_2(q^2 + c^2\om_n)$)
\equa{vp_1}
K_2 \left ( -\frac{\partial^2}{\partial z^2} + q^2 + g(z) \right) \varphi_{n}(z)=
\ep_{\om}\varphi_{n}(z)
\auqe
the solutions of which are known \cite{zitt, brez}; the spectrum of
eigenfunctions includes two bound states, $\varphi_0~=~C_0/\cosh^2(cz)$ and
$\varphi_1~=~C_1\sinh(cz)/\cosh^2(cz)$, with $\om_0$ = 0, $\om_1$ = 3 respectively
and a subset of unbounded states, or continuum
spectrum $\fik$, with $k$ = \mbox{$c\sqrt{\om_k-4}$} which behave as plane waves in the
bulk phases, {\it i.e.} far from the interface, and are given by
\equa{phi_k}
\fik = C_1(k)e^{ikz} \left[ 2 - \frac{k^2}{c^2} -\frac{3 i}{c}~k~\tanh(cz)
+ 3~(\tanh^2(cz) - 1) \right]
\auqe
It is important to note that 
\mbox{$\varphi_0(z) \propto \partial\phi_c(z)/\partial{z}$}.
Then the subset of unbounded eigenstates must be orthogonalized; using the eigenvalue
equation (\ref{vp_1}) and given the form of the $\fix{k}$ (\ref{phi_k}), we get a non
trivial dispersion relation leading to the density of states ($x = k/c$)
\equa{dos}
\rho(k) = \frac{L}{\pi} - \frac{1}{\pi c} \left[
\frac{1}{(1+x^2)} + \frac{2}{(4+x^2)} \right] =
\frac{L}{\pi} + \frac{f(x)}{c}
\auqe
where $L$ is the size of the system in the $z$ direction. It is important 
to note that \mbox{$\int_{-\infty}^{\infty} f(x) dx$ = -2} which proved useful in all 
calculations. Zittartz \cite{zitt} have already got this density of states
but without explicitely mentionning the need of the orthogonalization.
In \cite{dicap} a similar kind of dispersion relation was obtained in the
modelling of the charge density profile of electrolytes in the framework
of another field theory.
We emphasize that the orthogonalization of the eigenstates
is of crucial importance for the calculation of gaussian functional integrals.
In order to calculate $\ln(\Xi)$ and the correlation functions, we write
the fields $\chi(\vec{r})$ in the basis
which diagonalize the
operator $H^{(2)}$, namely a Fourier transform in the direction $\vec{s}$ and a
projection on the $\fix{\lambda}$ where $\{\lambda\}$ = $\{n=0,1; k\}$ denotes
the whole spectrum of eigenstates.
From usual gaussian functional integrals \cite{binney} we get
\equa{log_xi}
\ln(\Xi) = - H[\phi_c] +
\frac{V}{2} \int \frac{d_3\vec{k}}{(2\pi)^3}
~\ln\left(\frac{\hat{w}^{-1}(k)}{\ep_k(q)}\right)
\\ -
\frac{S}{2} \int \frac{d\vec{q}}{(2\pi)^2} \left(
\sum_{n}~\ln \left( \ep_{n}(q) \right)
+ \int dk f (k/c)~\ln \left( \ep_{k}(q) \right)
\right)
\nonumber \auqe
where $\ep_{k}$ corresponds to the second functional derivative of the
Hamiltonian for the $\phi^4$ homogeneous model taken at $\phi~=~\phi_c(\pm\;L)$.
Therefore, the volume term of $\ln(\Xi)$ in (\ref{log_xi}) is nothing but 
\mbox{$\be pV$} at the one loop level of approximation \cite{jmc_1}.
We finally write the surface term of
$\ln(\Xi)$ in term of the surface tension $\ga=[\ln(\Xi_b) - \ln(\Xi)]/S$ =
$\ga_{MF} + \ga^{(1)}$. We obtain ($\ga^{(1*)}$ = $\be\ga^{(1)}\s^2$)
\equa{gamma_1}
\ga^{(1*)} = \frac{1}{8\pi}
\left[ (x-c^2)\ln\left(1 - \frac{2c}{\sqrt{x}+c} \right)
\right. \nonumber \\ \left. +
 (x - 4c^2)\ln\left(1 - \frac{4c}{\sqrt{x} + 2c}\right)
- 6c\sqrt{x} \right]_{q_m^2+4c^2}^{q_M^2+4c^2}
\auqe
where $q_m$ and $q_M$ are the lower and upper bound respectively of the integral
over $q$. A similar result was obtained in \cite{zitt} for a spin model.
This result differs from that obtained in \cite{lipo_5}
where only the $n = 0$ eigenstate is kept.

We now consider the calculation of the field two-body correlation functions,
\linebreak
\mbox{$G(1,2) = <\chi(1)\chi(2)>_{H}$}.
More precisely, we focus on the Fourier transform parallel to the surface,
$G(z_1,z_2,q)$. For this we have to calculate the sum
\equa{corr1}
\sum_{\la}\frac{\varphi_{\la}(z_1) \varphi_{\la}^{*}(z_2)}{\ep_{\la}(q)} =
\sum_{n=0,1}\frac{\varphi_{n}(z_1) \varphi_{n}(z_2)}{\ep_{n}(q)}
+ \int\rho(k)\frac{\varphi_{k}(z_1) \varphi_{k}^{*}(z_2)}{\ep_{k}(q)} dk
\nonumber \auqe
A contribution of the integral over $k$ cancels exactely the direct contribution 
of the two bound states, which shows, once again, that the approximation 
consisting in keeping only the $\fix{0}$ eigenmode is not sufficient. The result 
is ($z_{>,<} = sup,inf(z_1,z_2)$)
\pagebreak[0]
\equa{corr_1}
G(z_1,z_2,q) = \nopagebreak[4]
\frac{9}{2cK_2}\frac{\exp(-c\abs{z_{12}}((q/c)^2+4)^{1/2}) }{((q/c)^2+4)^{1/2}(q/c)^2(((q/c)^2+3)}
\nonumber \\ \nopagebreak[4]
\textrm{x}\left[1+q^2/3c^2 + ((q/c)^2+4)^{1/2}\tanh(cz_>) + \tanh^2(cz_>) \right]
\nonumber \\ \nopagebreak[4]
\textrm{x}\left[1+q^2/3c^2 - ((q/c)^2+4)^{1/2}\tanh(cz_<) + \tanh^2(cz_<) \right]
\auqe
%
%%%%  Section 3 apres cette ligne
\section { Results and discussion}
\label{result}
\subsection*{Capillary behavior and surface structure factor}
We started from the expansion of the effective Hamiltonian on the basis of the 
eigenstates $\fix{\la}$. The first eigenstate, $\fix{0}$ is proportional to the
derivative of the mean field result $\phi_c(z)$, the proportionality constant 
being determined by the normalization. If we keep only this first eigenstate, 
the expansion of $\chi$ reads $\chi(\vec{s},z)~=~\xi(\vec{s})\fix{0}$. Hence the 
field takes the form $\phi(z) = \phi_c(z-z_{int}(\vec{s}))$ and only the linear 
term in the expansion of $\phi$ with respect to $z_{int}$ is kept. This 
corresponds to the so-called rigidly displaced profile approximation, where 
$z_{int} = -a\xi(\vec{s})$ represents the fluctuating location of the interface.
The corresponding contribution to $H$ is a functional of $\xi(\vec{s})$ which
defines an effective surface Hamiltonian given by
\equa{cwt_1}
H_s^{(0)}[\xi(\vec{s})] = \frac{1}{2} \int d\vec{s}
\left[ K_2(\partial_{\vec{s}}(\xi(\vec{s}))^2 + \om_0\xi(\vec{s})^2 \right]
\auqe
using $\om_0$ = 0, and the mean field
equation yielding $K_2/a^2$ = $\be\ga_{MF}$ we rewrite (\ref{cwt_1}) in the form
\equa{cwt_2}
H_s^{(0)} = \frac{1}{2} \be\ga_{MF} \int d\vec{s} (\partial_{\vec{s}}(z_{int}(\vec{s}))^2
= H_{CWT}
\auqe
which coincides with the usual effective surface Hamiltonian of the $CWT$ theory
for the free surface, {\it i.e.} in the absence of external field. Therefore we
obtain the $CWT$ as the lowest approximation beyond the mean field approximation
without invoking phenomenological arguments.

The structure is also characterized by the height-height correlation function,
$<z_{int}(\vec{s}_1)z_{int}(\vec{s}_2)>$ or its Fourier transform parallel to
the surface which defines the surface structure factor, $S_{\gD{z}\gD{z}}(q)$,
where $z_{int}(\vec{s})$ is the location of the interface relative to its mean
value. We consider  $\int\chi(\vec{s},z) dz/\gD\phi_b$, where
$\gD\phi_b = \phi_c(L)-\phi_c(-L)$, as a measure of the
instantaneous location of the surface at $\vec{s}$, which amounts to define
the location of the surface from a constraint on the integral of $\chi$,
as is done in \cite{blokhuis} for the density profile. We are
then led to identify $\szz(q)$ = \mbox{$(\gD\phi_c(b))^{-2}\int G(z_1,z_2,q)dz_1dz_2$},
which corresponds to the $S_{ic}$ used in \cite{blokhuis}.
It is important to notice that the coupling with the bulk fluctuations are included
in the present formulation through the eigenstates of the continuum.
We also introduce the effective surface width, or surface corrugation,
\mbox{$\s^{eff} = <z_{int}(\vec{s}_1)z_{int}(\vec{s}_1)>$}.

The behavior of $\szz(q)$ is analysed from the function
\mbox{$\mt{g}(q) = \int dz_1\int dz_2 G(z_1,z_2,q)$}.
From (\ref{corr_1}) we get for the leading term of $G(z_1,z_2,q)$ when \mbox{$q\tend 0$}
\equa{g_q_zero}
G(z_1,z_2,q \tend 0) ~\simeq~ \frac{1}{K_2 q^2} \varphi_0(z_1) \varphi_0(z_2)
\auqe
which corresponds to the $CWT$ behavior. Then it is easy to show that
\mbox{$1/(\gD\phi_b)^2\mt{g}(q \tend 0)$} $\tend 1/(\ga_{MF}q^2)$
from the relation already used between the normalization of $\fix{0}$ and $\ga_{MF}$
($(\int\fix{0}dz)^2$ = $(\gD\phi_b)^{2}K_2/\ga_{MF}$). This is exactly the
$CWT$ behavior, leading to the well known logarithmic divergence of the
surface corrugation for which we get
$\s^{eff}$ = \mbox{$(4\pi\ga_{MF})^{-1}\ln(q_{M}^2/q_{m}^2)$}.

\begin{figure}[h]
\begin{center}
\includegraphics[width = 0.6\textwidth]{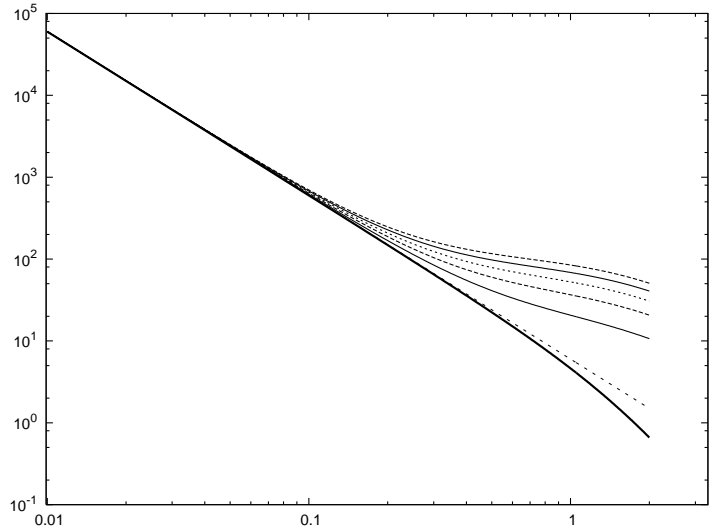}
\caption {\label{correl_int_bulk}
\protect\parbox[t] {0.6\textwidth}
{Log-log plot of $\mt{g}^*(x)$ versus $x~=~q/c$ for
% {$\ln(\mt{g}^*(x))$ versus $\ln(x)$ ($x$ = $q/c$) for
$Lc$ = 100, 80, 60, 40 and 20 from top; interfacial contribution
$\gD\phi_b^2S_{int}^*(x)$, bottom; $CWT$ limit, straight line.
}}
\end{center}
\end{figure}

It can be shown from (\ref{corr_1}) that the contribution to $\mt{g}(q)$ diverging
with the system size, due to the bulk correlations,
is exactly $(2L)$ times the integral of $G_b(z_{12},q)$,the correlation function
of the bulk phase, over $z_{12}$ If we keep only these two terms we get
\equa{corr_3}
\szz(q) \simeq
\frac{2L}{\gD\Phi^2}\int G_b(q,z_{12})dz_{12} + \frac{1}{\ga_{MF}q^2}
\auqe
Given that the bulk term vanishes when $q \tend 0$,
we see that (\ref{corr_3}) presents a cross-over like behavior in terms of
wave vector $q$, where a threshold value $q_s$ naturally appears,
separating the capillary wave behavior at small values of $q$ from the
bulk like behavior at $q > q_s$, with $q_s$ given by
\equa{qs}
q_s~\sim~[(2L\ga_{MF}/\gD\Phi^2)\int G_b(q,z_{12})dz_{12}]^{-1/2}.
\nonumber \auqe
Such a behavior is in agreement with that of Refs. \cite{blokhuis, sim_binder};
however, in the present formulation the bulk fluctuations come out of the 
calculation through the continuum subset of eigenstates and have not to be added
to the field profile. Furthermore, from (\ref{corr_3}) we can drop
exactly the bulk contribution, and doing this we define a purely
interfacial contribution to $\szz(q)$ (see fig.(\ref{correl_int_bulk})).
\equa{s_int_1}
S_{int}(q) =
\szz(q) - (2L/\gD\phi_b^2)\int G_b(q,z_{12})dz_{12}
\auqe
Then the departure of $S_{int}(q)$ from its $1/q^2$ behavior allows us to isolate
the deviations from the capillary wave like behavior of $\szz(q)$. To this aim we
define $\be(q)$ = \mbox{$(q^2S_{int}(q))^{-1} - \ga_{MF}$}. $\be(q)/q^2$ is found
nearly constant and thus lead us to define a bending rigidity of the interface,
\equa{kappa}
\kappa = \lim_{q\tend 0} ((1/q^2)\be(q)) \simeq ((1/q^2)\be(q))
\auqe
which takes a positive value, as it should be for the stability of the interface
\cite{dft_md}.
In other words, the small $q$ behavior of $S_{int}(q)$ is
\equa{s_int_2}
S_{int}(q) \simeq 1/[\ga_{MF} q^2 + \kappa q^4]
\nonumber \auqe
Here we find $\kappa$ = \mbox{$\kappa^*(\ga_{MF}\s^2)/(\s c)^2$}, with the
reduced value $\kappa^*$ = 0.288.
%
% \vskip 0.5 in

\subsection*{Surface tension}
\begin{figure}[H]
\begin{center}
\includegraphics [width = 0.6\textwidth]{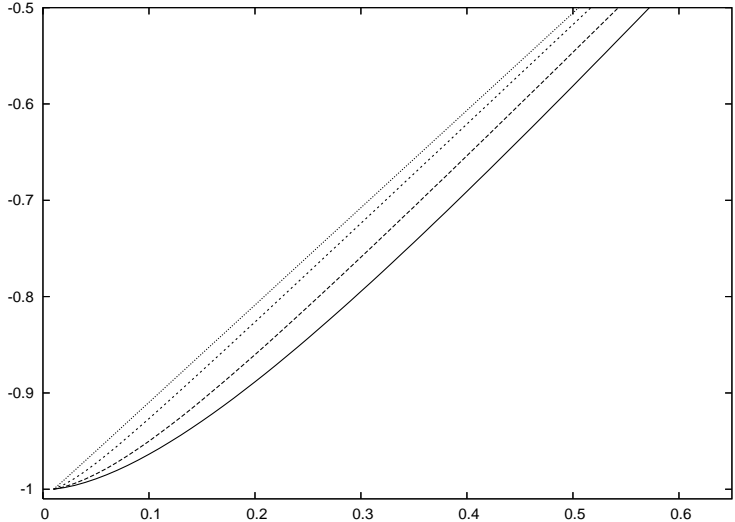}
\caption {\label{gamma1_red}
\protect\parbox[t] {0.6\textwidth}
{$\gamma^{(1)}(\sigma/L_x)/\gamma^{(1)}(0)$ in terms of $\sigma/L_x$
for \mbox{$q_M/(2\pi c)$ = 1}; 2; 5 and 50 from bottom to top.
}}
\end{center}
\end {figure}

The contribution $\ga^{(1)}$, given by (\ref{gamma_1}),
depends on the two bounds $q_m$ and $q_M$ of the wave vector $q$
which are related to the relevant parameters of the interface:
on the one hand 
$q_M/c$ = $2\pi/(\s{c})$ where $1/(\s{c})$ is the intrinsic width of the
interface in unit $\s$, $q_M/c$ $\in$ $[1,\infty[$. On the other hand,
$q_m/c$ = $(2\pi)(L_x/c)$ where $L_x$ is the system lateral size;
hence $q_m/c$ = $(q_M/c)(\s/L_x)$. Thus $q_M/c$ appears as a natural parameter,
with $q_M/c~\in~[2\pi, \infty]$ and $1/(q_M/c)~\propto~c~\propto~\sqrt{t}$.
We can now re-write $\ga^{(1*)}$ is a more convenient form
($\s/L_x~\in~]0,1]$)
\equa{gamma_1b}
\ga^{(1*)} = \frac{\pi}{2(q_M/c)^2} \left[ \mt{\ga}(4+(q_M/c)^2) -
         \mt{\ga}(4+(\frac{q_M}{c}\frac{\s}{L_x})^2) \right]
\auqe
$(L_x/\s)$ is either the actual lateral system size or the scale at which $\ga$ is
measured, for instance in a numerical simulation (see ref. \cite{tarazona}), 
but in any case does not depend on $t$. We can note that whatever
the values of $\s/L_x$ or $(q_M/c)$, $\s^2\ga^{(1)}$ remains finite and
more precisely
\equa{gamma_1lim}
\ga^{(1*)}_{L_x \tend \infty} (q_M/c \tend \infty) \sim -\frac{6\pi}{q_M/c}
\auqe
which $\tend$ 0 when $t \tend$ 0 as $t^{1/2}$.
The results for $\s^2\ga^{(1*)}$ are displayed in fig. (\ref{gamma1_red}).
We interpret $\s^2\ga^{(1*)}(\s/L_x)$ as the $q-$dependent contribution to $\ga$
with $q~=~(2\pi\s/L_x)$.
This means that the flat interface corresponding to the mean field approximation
is obtained when no fluctuation at all are taken into account, namely for 
\mbox{$q \sim q_M$}. This differs from what is done in Ref. \cite{dft_md} (see also ref. 
\cite{tara_comm}) where the contribution to $\ga$ due to the surface fluctuations
vanishes at \mbox{$q \tend 0$}.
The small $q$ behavior of $\ga^{(1*)}$ is easily obtained from (\ref{gamma_1b})
and yields
\equa{gamma1_lim2}
\ga^{(1*)}_{q\tend 0}
\simeq \ga^{(1*)}(0) + \frac{1}{8\pi}[(q\s)^2[a + \ln(c\s)] - 2(q\s)^2\ln(q\s)]
\makebox[0.1\textwidth]{with} a = 3.48491
\auqe
from which we can deduce a crossover value of $q$ given by $q_{0}\s$ =
$(c\s)e^{a/2}$,
separating the $q^2$ behavior from the $(q\s)^2\ln(q\s)$ behavior obtained for
\mbox{$q < q_{0}$} and \mbox{$q > q_{0}$} respectively. It is important
to note first that we always get an increasing $\ga^{(1)}(q)$ and
secondly that since $q_{0}$ is proportional to $c$,
we get a plateau (corresponding to the $q^2$ dependence) at small
values of $\s/L_x$ only when $q_M/c$ takes small values, $i.e.$ for the lowest
temperatures (see figure (\ref{gamma1_red}). This behavior is in qualitative 
agreement with the simulation results of \cite{tarazona}.

Notice that the term proportional to $q^2$ in the variation of $\ga^{(1)}$ 
with $q$ can be interpreted as resulting from the energy necessary to bend 
the interface, and should be related to $\kappa$ obtained from the behavior 
of $S_{int}(q)$ (see eq. (\ref{kappa}). This is not {\it a priori} the case 
since we do not expect a fully coherence in the framework of a loop 
expansion between the energetic and structure quantities.

\subsection*{Concluding remarks}
To conclude, we have shown in this work that the simple one-loop
expansion of the grand potential in the the simplest inhomogeneous situation
provides a qualitative but nevertheless coherent picture of the structure
of the liquid vapor interface of simple liquids. We emphasize that the
spectrum of eigenstates resulting from the diagonalization of the
second functional derivative of the effective Hamiltonian must be treated
as a whole and one cannot take into account only the first bounded state as
sometimes done in the literature.
Moreover, starting from a 3D Hamiltonian, and without invoking a 
phenomenological description of the interface, we recover the usual $CWT$
surface Hamiltonian as the first approximate step. We also show that 
we can extract a rigidity bending factor which takes a positive value,
in agreement with the requirement for the stability of the interface with 
respect to fluctuations. 
\eject
%%%%%%%%%%%%%%%%%%%%
%%%%% Biblio %%%%%%%
\begin{singlespace}

\end{singlespace}
\end{document}